\documentclass{PoS}
\title{Universality of the $N_f=2$ Running Coupling in the Sch\"odinger Functional Scheme.}

\ShortTitle{Universality of the $N_f=2$ Running Coupling}

\author{\speaker{Keiko Murano},  
Sinya Aoki\thanks{Riken BNL Research Center, Brookhaven National
Laboratory, Upton, NY 11973, USA}, 
Yusuke Taniguchi
 \\
        Graduate School of  Pure and Applied Sciences, University of Tsukuba, Tsukuba, Ibaraki 305-8571, Japan\\
        E-mail: \email{murano@het.ph.tsukuba.ac.jp} \\
        E-mail:\email{saoki@het.ph.tsukuba.ac.jp} \\
        E-mail:\email{tanigchi@het.ph.tsukuba.ac.jp}
        }

\author{Shinji Takeda
\\
 Humboldt Universitat zu Berlin
 Fachbereich Physik
 Inst. fur Elementarteilchenphysik
 Newtonstr. 15
 D-12489 Berlin, GERMANY\\
       E-mail: \email{takeda@physik.hu-berlin.de}}       
       
\abstract{We investigate universality of the $N_f=2$ running coupling in the Sch\"odinger functional scheme, by calculating the step scaling function in lattice QCD with the renormalization group (RG) improved gauge action at both weak($u=0.9796$) and strong($u=3.3340$) couplings, where
$u=\bar{g}^2_{\rm SF}$ with $\bar{g}_{\rm SF}$ being the running coupling in this scheme.
In our main calculations,  we use  the tree-level value for $O(a)$ improvement coefficients of boundary gauge fields. In addition we employ the 1-loop value for them in order to see how scaling behaviours are affected by them.  In the continuum limit,
the step scaling function obtained from the RG improved gauge
actions agrees with the previous result  obtained from the plaquette action within errors at both couplings, though errors of our result are larger.
Combined fits using all data with the  RG improved action as well as the plaquette action
reduce errors in the continuum limit by 2\% at the weak coupling and 22\% at the strong coupling.       
	  }

\FullConference{The XXVI International Symposium on Lattice Field Theory\\
		 July 14-19 2008\\
		 Williamsburg, Virginia, USA}

\begin{document}

\section{Introduction}

The $N_f=2$ $\beta$ function in the SF scheme has been 
calculated recently with the plaquette action and the $O(a)$
improved fermion action in Ref.\cite{DellaMorte:2004bc} and it is found that
the running coupling for $N_f=2$ QCD becomes stronger than that for the
pure gauge theory as the energy scale decreases. This behaviour is
opposed to the perturbative prediction and thus is a genuine
non-perturbative effect.
In such a non-perturbative region, however, it is well-known that the
calculation of the running coupling in the SF scheme becomes difficult, 
since the secondary minimum of the action comes close to the true
minimum so that  the auto-correlation time tends to be longer.
Therefore it is important to check the non-perturbative behaviour of the $\beta$ function mentioned above by using different gauge actions. 

In this study, employing the renormalization group (RG) improved gauge
action, we have calculated the step scaling function in the SF scheme
at both weak ($u=0.9793$) and strong ($u=3.3340$) coupling regions, 
where $u = \bar{g}_{\rm SF}^2$ with $\bar{g}_{\rm SF}$ being the running coupling
in the SF scheme.

\section{Set up}
Using the similar setup given in Ref.\cite{DellaMorte:2004bc}, 
we consider the improved gauge action on an $L^3\times T$ lattice 
in the SF scheme ,
\begin{eqnarray}
S_{imp} = \frac{2}{g^2} \times \left[\sum_{x}\omega_{\mu,\nu}^P(x_0)
\rm{Re} \rm{Tr}\left(1-P_{\mu,\nu}(x)\right)+\sum_{x}\omega_{\mu,\nu}^R(x_0)
\rm{Re}\rm{Tr}\left(1-R^{(1\times 2)}_{\mu,\nu}(x)\right)\right],
\end{eqnarray}
where weight factors are given by
  \begin{eqnarray}
    \begin{array}{ll}
 
   \omega^P_{\mu,\nu}(x_0) = \left \{
			      \begin{array}{ll}
			       c_0 c_s (g_0^2) & x_0=0,T, \ \mu,\nu \ne 0 \\
			       c_0 c_t^P (g_0^2) & x_0=0,T-a, \ \mu = 0 \ \rm{ or } \ \nu= 0 \\
			       c_0 & \rm{otherwise,}
			      \end{array}
					\right.

 &

   \omega^R_{\mu,\nu}(x_0) = \left \{
			      \begin{array}{ll}
			       0 & x_0=0,T \ \mu,\nu \ne 0 \\
			       c_1 c_t^R (g_0^2) &x_0=0,T-a \  \nu= 0 \\
			       c_1 & x_0=0,T, \mu = 0 \\
			       c_1 & \rm{otherwise,}
			      \end{array}
					\right.

    \end{array}\nonumber
  \end{eqnarray}
for a plaquette $P_{\mu,\nu}$ and an $1\times 2$ rectangular
$R^{(1\times 2)}_{\mu,\nu}$ with the constraint that $c_0 + 8 c_1 =
1$. 
For the RG improved gauge action, we take $c_0 = -0.331$. 
To removed $O(a)$ scaling violations caused by boundaries, we have to tune the
$O(a)$ improvement coefficients $c_t^R$ and $c_t^P$, which have been  calculated perturbatively at 1-loop\cite{Takeda:conditionA} as
  \begin{eqnarray}
   c_0c_t^P(g_0^2) &=& c_0 (1 + c_t^{P(1)}g_0^2 + O(g_0^4) )\nonumber \\
   c_1c_t^R(g_0^2) &=& c_1 (3/2 + c_t^{R(1)}g_0^2 + O(g_0^4) ) . 
  \end{eqnarray}
  The scaling study of the SF running coupling for the SU(3) pure gauge
  theory with this gauge action\cite{Takeda:2004xh}, however,  showed
  that the scaling violation at the strong coupling becomes larger for
  the 1-loop value of
  $c_t^{P.R}$ than that for the tree-level value. In this study
 we therefore take the tree-level value, $c_t^P = 1$ and $c_t^R=3/2$.
We have also performed an additional set of simulations with the
 1-loop value of $c_t^{P,R}$, in order to see how scaling behaviour are
  affected by the choice of $c_t^{P,R}$.
For quarks,
we employ the $O(a)$ improved Wilson fermion action (clover action) in
the SF scheme\cite{DellaMorte:2004bc}, with the non-perturbative value of
the improvement coefficient $C_{sw}$\cite{Aoki:2005et} 
and  the 1-loop  value of the improvement coefficient
$\tilde{c}_t$\cite{Aoki:1998qd}.

\section{Simulation details}
We have calculated the step scaling function (SSF) in the weak coupling region
($u=0.9793$) and the strong coupling region ($u=3.3340$).
Following the calculation procedure and the analysis in
Ref.\cite{DellaMorte:2004bc},
we have calculated  both $\bar{g}^2_{\rm SF}(L,a/L)$ and $\bar{g}^2_{\rm SF}(2L,a/L)$
at the same $\beta$, in order to obtain the lattice SSF as
\begin{eqnarray}
\Sigma(u, a/L) &=& \bar{g}^2_{\rm SF}(2L,a/L), \quad u=\bar{g}^2_{\rm SF}(L,a/L) .
\end{eqnarray}
In order to make the continuum extrapolation as $\displaystyle\lim_{a/L\rightarrow0}\Sigma(u,a/L)=\sigma(u)$, we repeat this procedure by
changing $L/a$ and $\beta$
while keeping $u=\bar{g}^2_{\rm SF}(L,a/L)$ fixed.
Throughout calculations, we tune the hopping parameter $\kappa$ so that 
the PCAC quark mass $m(x_0)$ at $x_0=T/2$ becomes zero at given $L/a$
and $g_0$, where
\begin{eqnarray}
   m(x_0) = \frac{\frac{1}{2}(\partial_0 + \partial_0^*)f_A(x_0) + c_A
    a\partial_0^*\partial_0 f_P(x_0)}{2f_P(x_0)} .
\end{eqnarray}
Throughout this study we take $T=L$.

Errors of the SSF $\Sigma$ due to the small deviation of $u$ from 0.9793
or 3.3340 are perturbatively corrected as
\begin{eqnarray}
  \Sigma(u,a/L) &=& \Sigma(\tilde{u},a/L) + \Sigma'(u,a/L) \times
    (u-\tilde{u}).  \\
   \Sigma'(u,a/L) &=& \frac{\partial \Sigma(u,a/L)}{\partial u} \sim
    \frac{\partial \sigma}{\partial u} \sim 
    1 + 2s_0 u + 3s_1 u^2 + 4s_2 u^3 .
\end{eqnarray}
Similarly, errors of the SSF due to the small non-zero value of the PCAC quark mass are corrected as
\begin{eqnarray}
&& \Sigma(u,a/L,z) = \Sigma(u,a/L,0) +
  \left.\frac{\partial}{\partial z} \Sigma(u,a/L,z)\right|_{z=0} \times z ,\quad z=m(L/2)L.\label{eq:PTmass0}\\
&& \left.\frac{\partial}{\partial z} \Sigma(u,a/L,z)\right|_{z=0} \sim
 \left.\frac{\partial}{\partial z} \sigma(u,a/L,z)\right|_{z=0} = 0.00957N_f u^2.\label{eq:PTmass}
\end{eqnarray}
where the right-hand side of eq.(\ref{eq:PTmass}) is taken from
\cite{Sint:1995ch}. In eq(\ref{eq:PTmass0},\ref{eq:PTmass}), we make a $z$ dependence of
$\Sigma$ explicit,  though hereafter  we denote it as $\Sigma(u,a/L)$ instead of $\Sigma(u,a/L,z)$  for 
simplicity.

We have carried out computations of the SSF at $L/a =4,6,8$ and $2L/a =8,12,16$. 
Details of simulation parameters and analysis will be given in Ref.\cite{MATT}.
Numerical simulations are performed on a cluster machine ``kaede'' at 
Academic Computing \& Communications Center, University of Tsukuba.

\section{Some remarks for numerical simulations}
In the previous study with the plaquette action\cite{Luscher:1993gh,DellaMorte:2004bc},
it has been reported that the auto-correlation time of the HMC updating tends to be longer 
in the strong coupling region. This long auto-correlation seems to be
caused by rare but large fluctuations of the gauge part of
$1/\bar{g}^2_{\rm SF}$, which appear as a result of  the tunneling
between the true minimum and the secondary minimum.
In order to make the auto-correlation shorter,
the modified gauge force which enhances a rate of such tunnelings
has been introduced  together with the reweighting method\cite{Luscher:1993gh,DellaMorte:2004bc}.

\begin{figure}[tbh]
\begin{center}
\scalebox{0.7}{\includegraphics{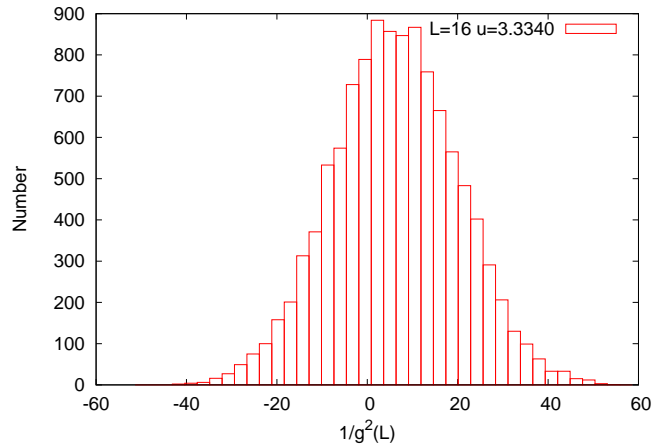}}
\end{center}
\caption{A distribution of the gauge part  of $1/\bar{g}^2_{\rm SF}$ with
 the RG improved action, in the case of 
 $L/a=16$, $\beta=2.755$ and $\kappa=0.13334$.}
\label{Fig:destribution}
\end{figure}

 We have checked whether a similar problem exists in the case of the RG
 improved action, by examining distributions of the gauge part of 
$1/\bar{g}^2_{\rm SF}$.
Fortunately we do not observe such rare but large fluctuations in distributions, and an example of distributions is shown in Fig.\ref{Fig:destribution}.
  We however observe that the distribution of $1/\bar{g}^2_{\rm SF}$
 with the RG improved action has much wider
 width than that  with the plaquette action\cite{private}.
 The wider width indicates that, while the HMC with the RG improved action
 samples configurations including  ones near the secondary minimum
 more effectively, statistical fluctuations  of $1/\bar{g}^2_{\rm SF}$ with the RG action 
become also larger than those with the plaquette action.

\section{Result}
\begin{figure}[tb]
 \begin{tabular}{cc}
  \scalebox{0.6}{\includegraphics{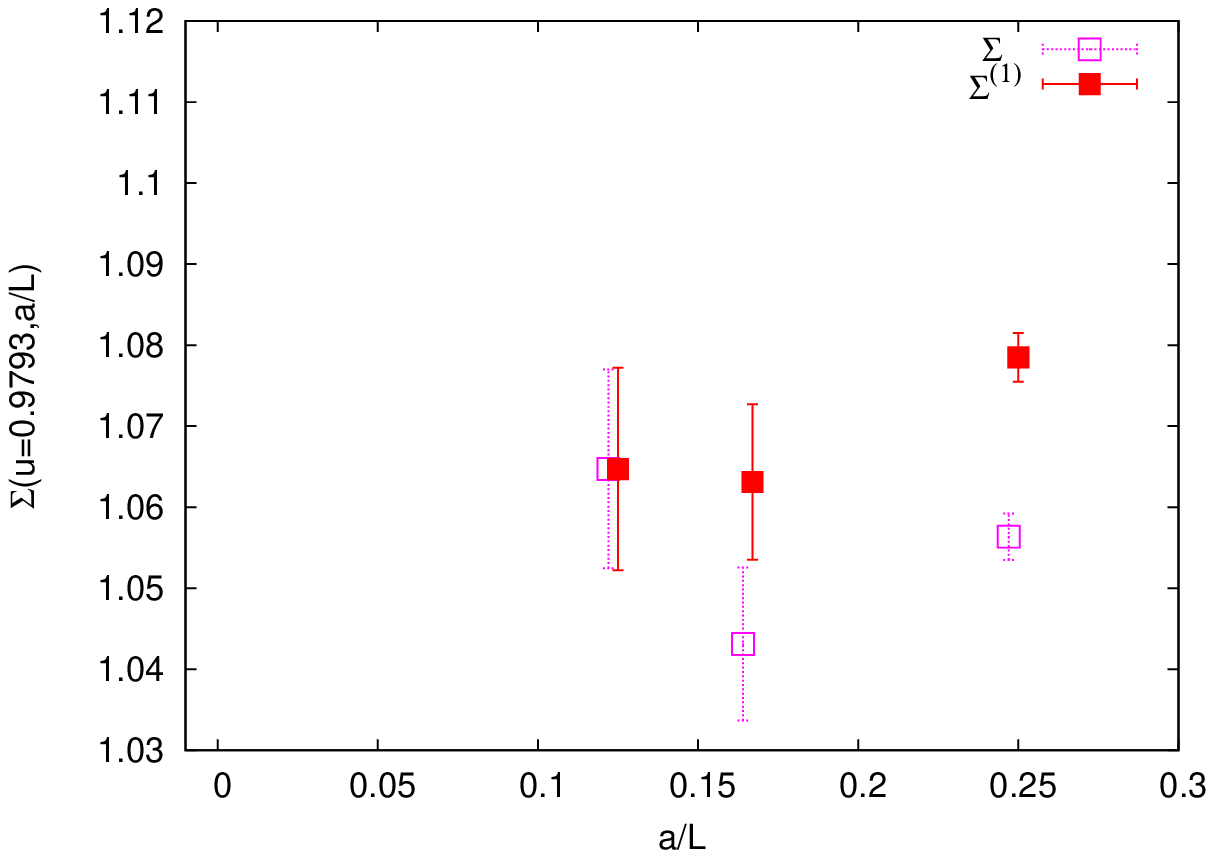}} &
  \scalebox{0.6}{\includegraphics{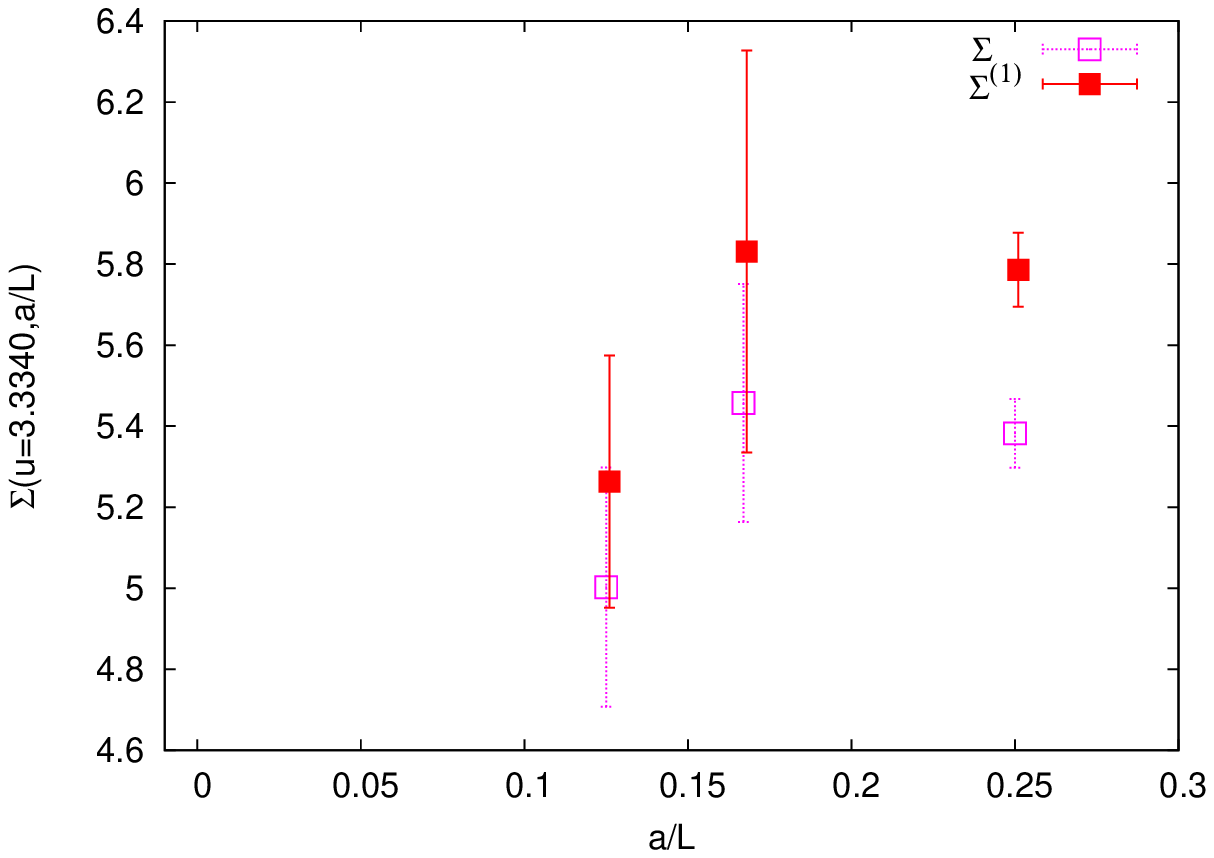}}
 \end{tabular}
\caption{Scaling behaviours of $\Sigma$(open squares) and
 $\Sigma^{(1)}$(solid squares)  at the weak coupling ($u=0.9793$, left)
and at the strong coupling ($u=3.3340$, right).}
\label{Fig:PTimp}
\end{figure}
In Fig.\ref{Fig:PTimp}, we compare scaling behaviours between 
the (naive) lattice SSF $\Sigma(u,a/L)$ and  the 1-loop improved one $\Sigma^{(1)}(u,a/L)$ for the RG-improved gauge action with the
tree-level value of $c_t$ at the weak coupling ($u=0.9793$, left) and the strong coupling ($u=3.3340$, right).
Here $\Sigma^{(k)}(u,a/L)$, whose lattice artifacts are perturbatively removed at k-loop,  is defined by 
  \begin{eqnarray}
   \Sigma^{(k)}(u,a/L) = \frac{\Sigma(2,u,a/L)}{1+\delta_1(a/L)u+\delta_2(a/L)u^2+\ldots +\delta_k(a/L)u^k}.
  \end{eqnarray}
   where $\delta_n$ $(n=1,2\ldots)$ is given by  
   \begin{eqnarray}
    \frac{\Sigma(u,a/L)-\sigma(u,a/L)}{\sigma(u,a/L)} 
     = \delta_1(a/L)u + \delta_2(a/L)u^2 + \ldots  ,
   \end{eqnarray}
and $\delta_1(a/L)$ for the RG improved action is given in Ref.\cite{Sint:1995ch,Takeda:2004xh}.
\begin{figure}[bt]
\begin{tabular}{cc}
\scalebox{0.6}{\includegraphics{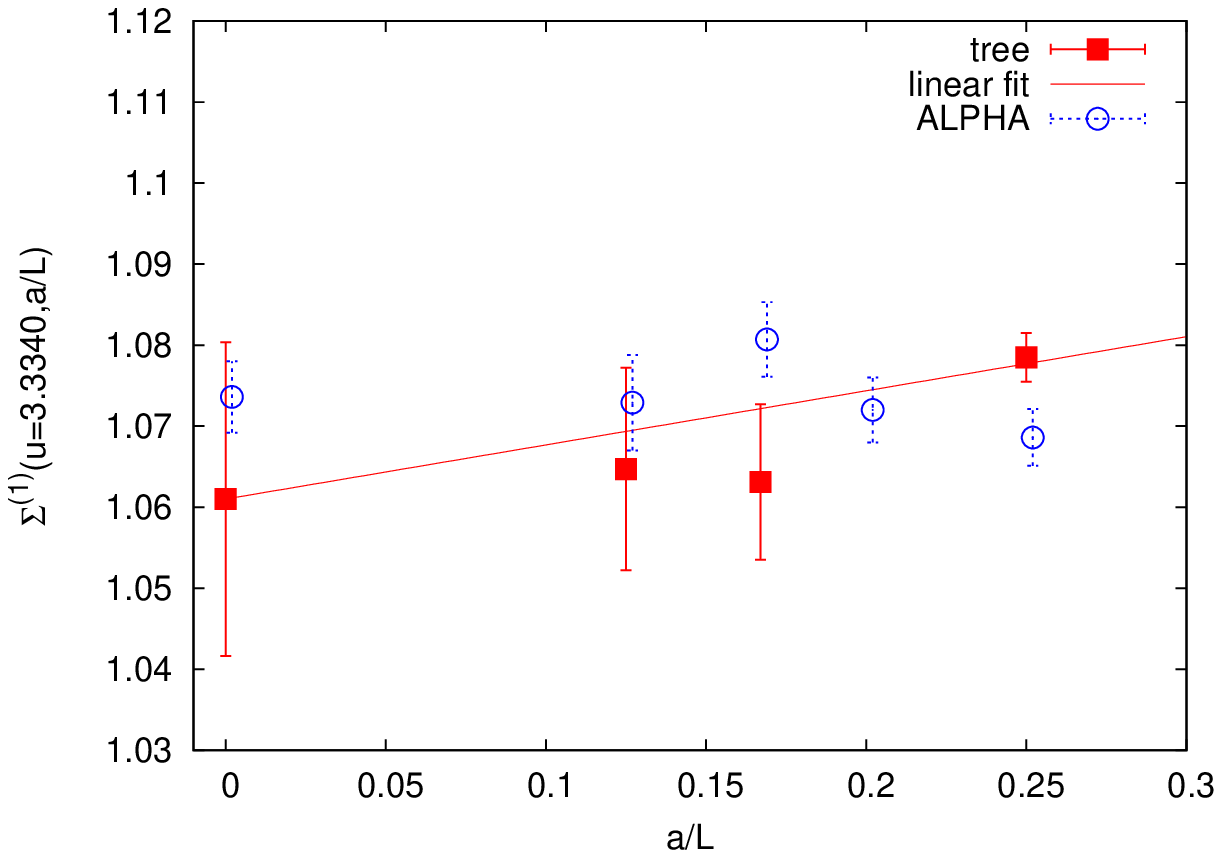}}
&
\scalebox{0.6}{\includegraphics{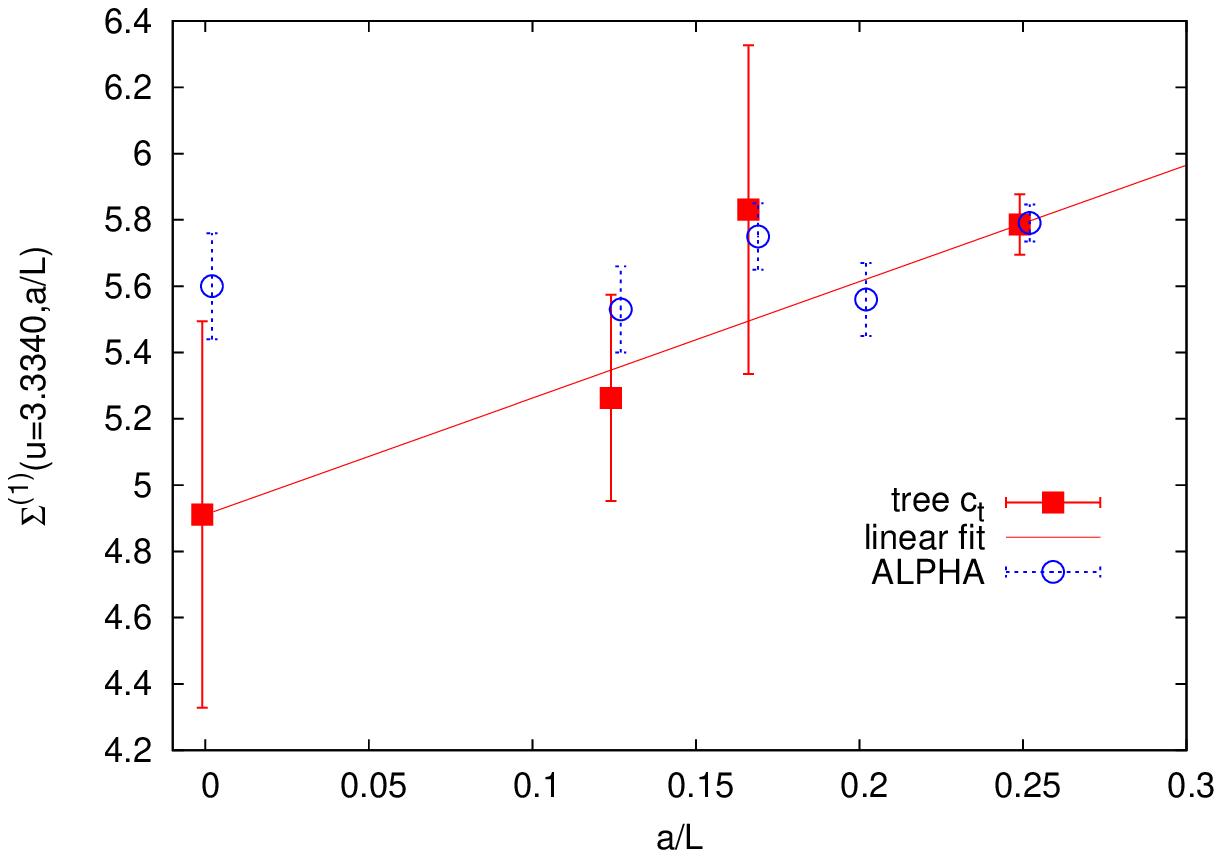}}
\end{tabular}
\caption{Linear continuum extrapolations of $\Sigma^{(1)}(u,a/L)$
at $u=0.9793$ (left) and at $u=3.3340$ (right),
together with 
the result of the ALPHA Collaboration\protect\cite{DellaMorte:2004bc},}
\label{Fig:continuumtree}
\end{figure}
Since the scaling violation of $\Sigma^{(1)}$ is milder than that of $\Sigma$,
in particular at $u=0.9793$,  
we hereafter use $\Sigma^{(1)}$ for the continuum extrapolation.

In Fig.\ref{Fig:continuumtree}, we compare our  $\Sigma^{(1)}$ with $\Sigma^{(2)}$ from the 
ALPHA Collaboration, where $\Sigma^{(2)}$  is the 2-loop improved lattice SSF, 
 and observe that our errors of $\Sigma^{(1)}$ are larger in general.
We think that these larger errors are caused partly by lower statistics and partly by wider widths of distributions mentioned in the previous section.
In the figure we also plot linear continuum extrapolations of $\Sigma^{(1)}$, which
give $\sigma(u=0.9793)=1.061(19)$ with $\chi^2/{\rm dof}=1.74$ and 
$\sigma(3.3340)=4.91(58)$ with $\chi^2/{\rm dof}=0.585$.
These values are consistent with results from the ALPHA Collaboration, $\sigma(u=0.9793)=1.072(4)$ and $\sigma(3.3340)=5.60(16)$,  though our errors are much lager as expected from errors of $\Sigma^{(1)}$.
Note however that results from the ALPHA Collaboration are obtained from the combined fit with  $\Sigma^{(2)}$ at all $u$ by the form $\sigma(u) + \rho^{\rm X-loop} u^4 (a/L)^2$
neglecting a tiny $a/L$ term, 
therefore errors tend to be smaller than those form the individual fit. 
Here 
$X$ is the order of the improvement coefficient $c_t$: 1-loop for  data at $u$ in the weak coupling region and 2-loop in the strong coupling region\cite{DellaMorte:2004bc}, and the scaling violation starts at 
$u^3$ (3-loop) in $\Sigma^{(2)}(u)/u$.

\begin{figure}[bt]
\begin{tabular}{cc}
\scalebox{0.6}{\includegraphics{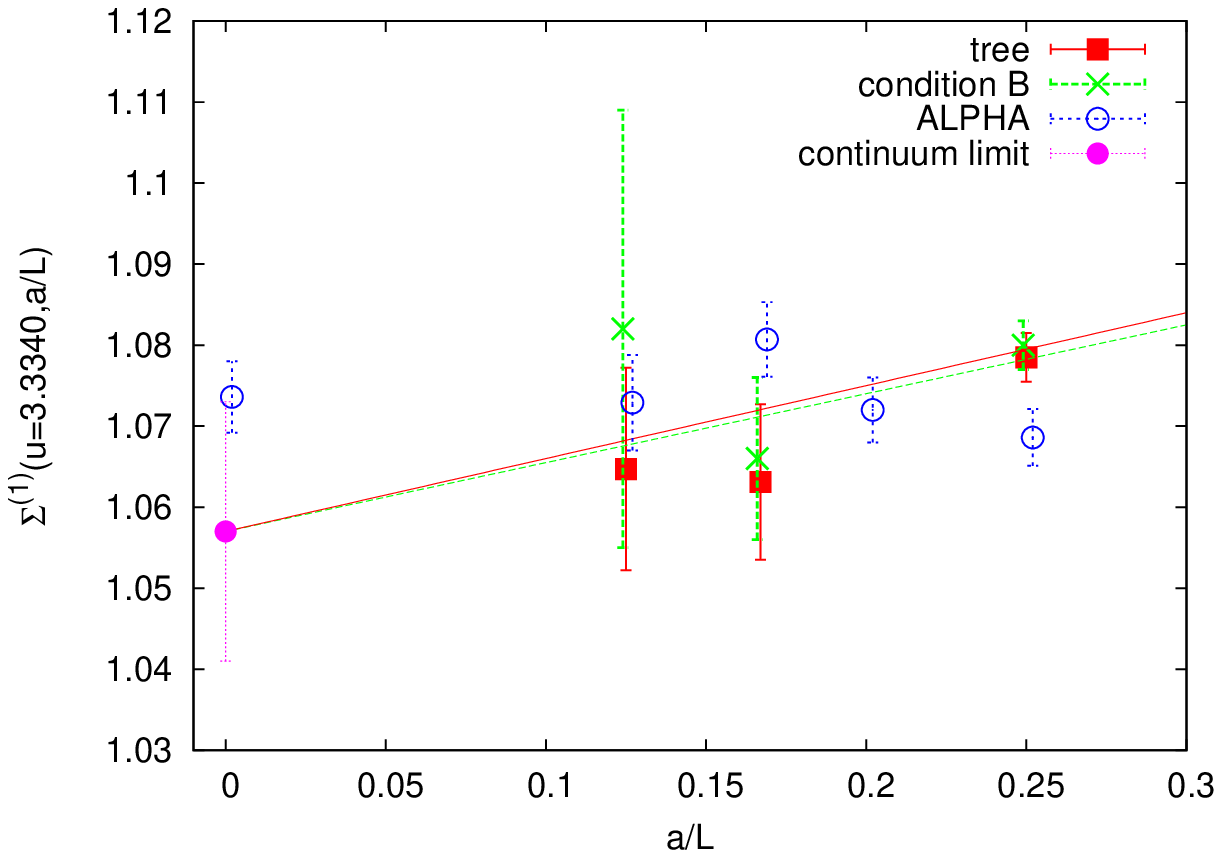}}
&
\scalebox{0.6}{\includegraphics{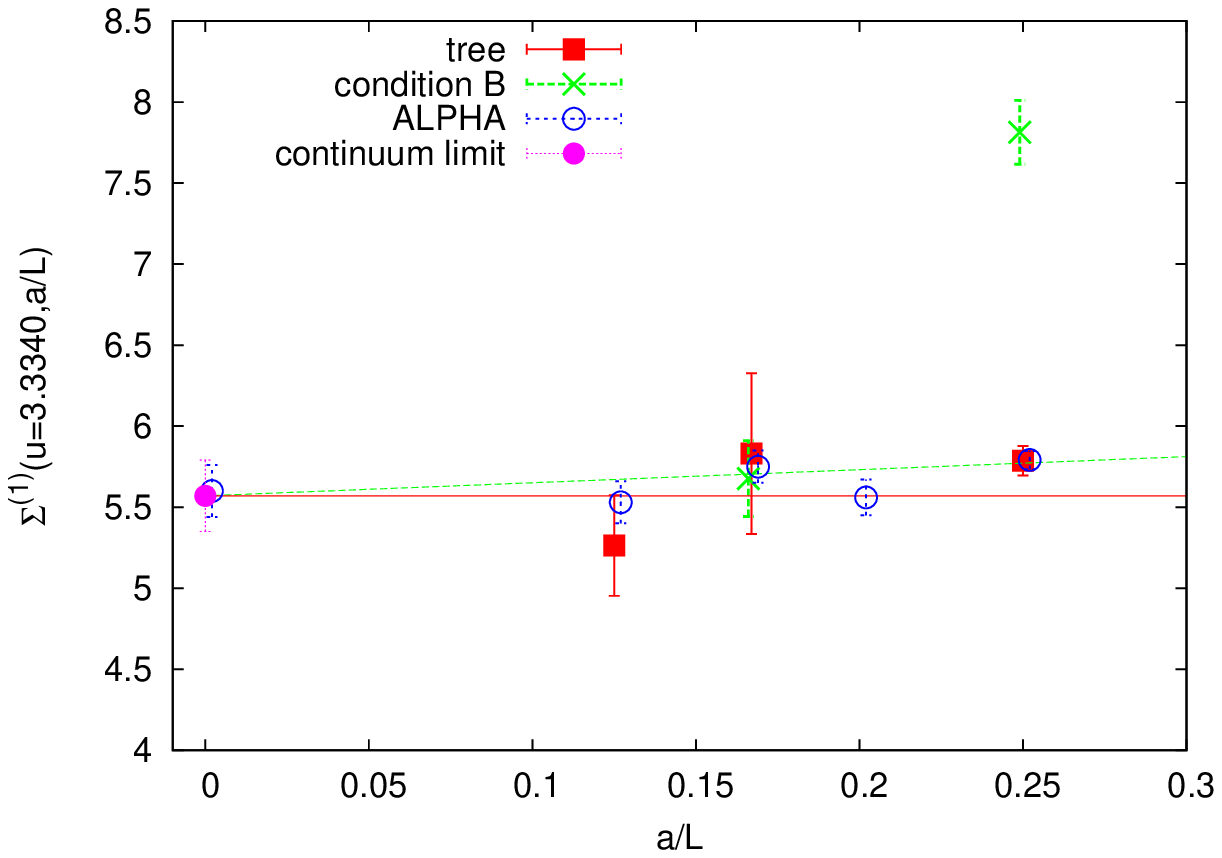}}
\end{tabular}
\caption{Combined fits to data including
the tree $c_t$ (squire) at all $L/a$ and the 1-loop $c_t$ (cross) except $L/a=4$,
at $u=0.9793$(left) and at $u=3.3340$(right).
Results from the ALPHA Collaboration are also plotted(open circles).}  
\label{Fig:continuumchoiceB}
\end{figure}

In order to examine scaling behaviours of $\Sigma^{(1)}$ with the RG improved action more precisely,
we have repeated the calculation of the SSF using the 1-loop value of boundary improvement coefficients $c_t^{P,R}$ with condition B\cite{Takeda:conditionA}.
$\Sigma^{(1)}$ from the 1-loop $c_t^{P,R}$ is plotted in Fig.\ref{Fig:continuumchoiceB},
together with results from the tree $c_t$ and from the ALPHA Collaboration\cite{DellaMorte:2004bc}.
We observe that three results show  mild scaling violations for all $a/L$s'  at the weak coupling(the left figure) , while, in the strong coupling region (the right figure), the scale violation of the result from the 1-loop $c_t^{P,R}$ becomes  too large at $L/a=4$ to make the  reliable continuum extrapolation with this point.
A possible reason for this large scaling violation is that the perturbative estimate of $c_t^{P,R}$
becomes unreliable in the strong coupling region of the RG-improved action due to the large value of the bare gauge coupling $g_0^2$.
The $\beta$ value which gives the same $u$ is much smaller  for the RG-improved action than
for the plaquette action. For example, at $u = 3.3340$ and $L/a = 4$,
the RG-improved action needs $\beta = 2.1361$ , which corresponds to $g_0^2=2.8089$ and gives a large(10\%) 1-loop correction to $c_t^R$, $c_t^{R(1)} g_0^2 = 0.146$.
On the other hand, the tree-level value, $c_t^R$=1.5, remains
unchanged for all $\beta$. 
Since $\bar{g}^2_{\rm SF}$ is defined through boundary observables, a small
difference in $c_t^{P,R}$ may have a large effect on it.

We now perform a combined fit using all our data with tree-level and
1-loop values of  $c_t^{P,R}$, except one with the 1-loop $c_t^{P,R}$ at $L/a=4$ and $u=3.334$.
At the  weak coupling( $u=0.9793$), we use linear functions in $L/a$ for data with
the tree-level value of $c_t^{P,R}$( 3 data) and the 1-loop values of $c_t^{P,R}$(3 data), while 
at the strong coupling($u=3.3340$), we use a linear fit for data with
the tree-level value of $c_t$( 3 data) and a constant fit for data with the 1-loop value of $c_t$ excluding one at $L/a = 4$( 1 data).
From the simultaneous fits shown in Fig. \ref{Fig:continuumchoiceB},
we obtain $\sigma(u=0.9793) = 1.057(16)$ with $\chi^2/{\rm dof}=1.22$ and $\sigma(u=3.334)=5.57(22)$ with $\chi^2/{\rm dof}=2.02$.
While both central value and error are almost unchanged at the weak coupling, 
an agreement with the result from the ALPHA Collaboration becomes better
with smaller error at the strong coupling.

We finally make a more complicated combined fit, using our data and fitting functions mentioned above
and  employing  the form $\sigma(u) + \rho^{\rm X-loop} u^4 (a/L)^2$ as fitting functions to data at all $u$ from the  ALPHA Collaboration\cite{DellaMorte:2004bc}.
We obtain $\sigma(u=0.9793) = 1.0724(43)$ and
$\sigma(u=3.334)=5.559(125)$ with $\chi^2/\rm{dof}=2.39$
in the continuum limit. 
Compared with result from ALPHA Collaboration, 
1.0736(44) for weak coupling and 5.60(16) for strong coupling, these errors are reduced by
2\% and 22\%,  respectively.

\section{Conclusion}
We have calculated the step scaling function in the SF scheme
at weak ($u=0.9793$) and strong ($u=3.3340$) coupling regions with
the renormalization group (RG) improved gauge action.
Extrapolated values of the SSF  from the RG improved gauge action
agree with those from  the plaquette action within errors at both couplings,
though errors of the former are larger.

From a combine fit using all data including ones
with the plaquette action from the ALPHA Collaboration\cite{DellaMorte:2004bc}, we obtain 
 $\sigma(u=0.9793) = 1.0724(43)$ and $\sigma(u=3.334)=5.559(125)$
in the continuum limit. These errors are reduced by 2\% and 22\%  from
previous results in Ref.\cite{DellaMorte:2004bc}.

Finally we make two comments on calculations of the SSF with the RG improved gauge action.
Firstly, it is better to use the tree-level value of  $O(a)$ improvement coefficients $c_t^{P,R}$ than the 1-loop value, in particular, in the strong coupling region.
Secondly, the HMC algorithm with the RG improved action samples configurations including
ones near the secondary minimum better than that with the plaquette action, though
statistical fluctuations  of $1/\bar{g}^2_{\rm SF}$ with the RG improved action become larger.

\section*{Acknowledgments}
We are grateful for authors and maintainers of {\tt CPS++}\cite{cps},
of which a modified version is used for simulations done in this work.
This work was supported in part by the Grant-in-Aid of the Japanese Ministry of Education, Culture, Sports, Science and Technology (Nos. 18740130, 20340047).

\end{document}